\documentclass[11pt,twocolumn]{article}
\usepackage{amsmath,amsfonts,amssymb}
\usepackage{graphicx,url,array,booktabs, multirow}
\usepackage[left=1.7cm,right=1.7cm,bottom=2.4cm,top=2.1cm]{geometry}
\usepackage{flushend,caption}
\title{{VOIC\MakeLowercase{e}}: A Sound Event Detection Dataset For Generalizable Domain Adaptation}
\author{Shayan Gharib, Konstantinos Drossos, Eemi Fagerlund, and Tuomas Virtanen\vspace{6pt}\\
\{firstname.lastname\}@tuni.fi\\
Audio Research Group, Tampere University, Finland}
\date{}
\begin{document}
\twocolumn[
\maketitle
  \begin{@twocolumnfalse}
    \maketitle
        \begin{abstract}
        The performance of sound event detection methods can significantly degrade when they are used in unseen conditions (e.g. recording devices, ambient noise). Domain adaptation is a promising way to tackle this problem. In this paper, we present VOICe, the first dataset for the development and evaluation of domain adaptation methods for sound event detection. VOICe consists of mixtures with three different sound events (``baby crying'', ``glass breaking'', and ``gunshot''), which are over-imposed over three different categories of acoustic scenes: vehicle, outdoors, and indoors. Moreover, the mixtures are also offered without any background noise. VOICe is freely available online\footnotemark. In addition, using an adversarial-based training method, we evaluate the performance of a domain adaptation method on VOICe.
        
        \vspace{6pt}
        \end{abstract}
        \vspace{18pt}
  \end{@twocolumnfalse}
]
\footnotetext{\url{https://doi.org/10.5281/zenodo.3514950}}
\section{Introduction}
\label{sec:intro}
Sound event detection (SED) is the task of identifying the onset and offset times of sound event classes (e.g. ``footsteps'', ``car horn'', and ``bird singing'') in an audio signal. In an audio sample, there can be multiple events at the same time, but monophonic SED systems detect only one event at any given time instance, while polyphonic SED aims at detecting all the predetermined target sound event classes~\cite{Virtanen2018, 7933050}. Applications of SED methods include, for example, automatic audio indexing~\cite{1621215}, surveillance~\cite{1521503}, and health care~\cite{5202720}. Initial SED methods were based on Gaussian mixture models, hidden Markov models, and non-negative matrix factorization~\cite{10.1007/978-3-540-68585-2_33, Dessein2013}. However, with the recent availability of bigger datasets for SED, deep neural networks (DNNs) are employed almost exclusively. For example, in the DCASE workshop and challenge\footnote{\url{http://dcase.community/}}, the recent state-of-the-art results are obtained by applying stacked convolutional and recurrent neural networks (CRNNs)~\cite{7933050,7760424}. The typical scenario of developing an SED method is firstly to optimize the method using audio signals $\mathbf{x}_S\sim\mathcal{X}_S$ and their labels by supervised learning. Then, we use the optimized method on different audio signals, $\mathbf{x}_T\sim\mathcal{X}_T$, where $\mathcal{X}_S$ and $\mathcal{X}_T$ are the underlying distributions of the audio signals.

One of the main problems of SED methods is the degradation of performance when there is a mismatch between $\mathcal{X}_S$ and $\mathcal{X}_T$. This mismatch might be a result of capturing the data under different conditions, for example different acoustic scenes, different recording devices, to name a few. This problem of data/datasets captured under different conditions is a general machine learning problem and is referred to as the capturing bias~\cite{Quionero-Candela:2009:DSM:1462129, Ben-David2010}. As a subset of the domain shift problem, the capturing bias degrades the performance of methods. Domain adaptation has been one of the active research areas to address the above mentioned issue of capturing bias, and domain shift in general~\cite{TTommasi_etal_GCPR2015, Torralba:2011:ULD:2191740.2191773}, in many applications such as machine vision~\cite{pmlr-v37-ganin15, 8099799}, speech recognition~\cite{6817520, SUN201779}, and audio classification~\cite{drossos2019unsupervised}. 
The general framework of existing domain adaptation methods is such that the parameters of method $\Theta_{S}$ are optimized using a set of data originating from a distribution $\mathcal{Z}_{S}$. Then, the hyper-parameters $\Psi$ (e.g. the architecture of a DNN, epochs of training, different weights of losses) of an adaptation process $f$ are fine tuned for adapting $\Theta_{S}$ to data coming from a different distribution $\mathcal{Z}_{T}$, as
\begin{equation}
    f_\Psi:\Theta_{S}\mapsto\Theta_{T}\text{,}
\end{equation}
\noindent
where $\Theta_{T}$ are the parameters of the method, adapted to $\mathcal{Z}_{T}$. Finally, the adaptation process $f_\Psi$ is assessed by measuring the performance of the adapted method (i.e. $\Theta_{T}$) on the data coming from $\mathcal{Z}_{T}$. Therefore, the results of the adaptation process are based on the same data used to fine tune the hyper-parameters of the adaptation process. This procedure does not guarantee the generalization of the adaptation process $f$: it is unknown if the same process $f_{\Psi}$ could be effective on adapting $\Theta_{S}$ on data coming from a third distribution $\mathcal{Z}_{T'}$. 

\begin{table*}[!t]
\centering
\caption{Three categories of acoustic scenes available in TAU Urban Acoustic Scenes 2019 dataset.}
\label{tab:table1}
\bigskip
\resizebox{\textwidth}{!}{%
\begin{tabular}{l|ccc}
\textbf{Category} & \textit{Vehicle} & \textit{Outdoor} & \textit{Indoor} \\ \hline
\multirow{4}{*}{\textbf{acoustic scenes}} & \multicolumn{1}{c|}{\multirow{4}{*}{\textit{\begin{tabular}[c]{@{}c@{}}travelling by a bus\\travelling by a tram \\ travelling by an underground metro\end{tabular}}}} & \multicolumn{1}{c|}{\multirow{4}{*}{\textit{\begin{tabular}[c]{@{}c@{}}urban park\\ public square\\pedestrian street \\ street with medium level of traffic\end{tabular}}}} & \multirow{4}{*}{\textit{\begin{tabular}[c]{@{}c@{}}airport\\ metro station \\ indoor shopping mall
\end{tabular}}} \\
 & \multicolumn{1}{c|}{} & \multicolumn{1}{c|}{} &  \\
 & \multicolumn{1}{c|}{} & \multicolumn{1}{c|}{} &  \\
 & \multicolumn{1}{c|}{} & \multicolumn{1}{c|}{} & 
\end{tabular}
}
\end{table*}

In this paper, we introduce VOICe, the first and publicly available SED dataset for generalizable domain adaptation. VOICe provides audio data and annotations for polyphonic SED, having the starting and ending time-stamps from three different sound events (namely baby crying, glass breaking, and gunshot) over three different categories of acoustic scenes (namely vehicle, outdoor, and indoor). Sound events and acoustic scenes are from also publicly available datasets, namely the TUT Rare Sound Events 2017 and TAU Urban Acoustic Scenes 2019, respectively. By employing three categories of acoustic scenes, VOICe allows to overcome the shortcoming of assessing the performance of the adaptation process on data coming from the same distribution as those used for fine-tuning the hyper-parameters of the adaptation process. The rest of the paper is organized as follows. In Section 2, we introduce the SED domain adaptation dataset. Section 3 presents the settings used for our experiments in addition to the baseline method and evaluation results. Finally, in Section 4, we describe the conclusion of our work.

\section{VOICe dataset}
\label{sec:dataset}
VOICe is an artificially created dataset using sound events from TUT Rare Sound Events 2017 development set~\cite{DCASE2017challenge} and acoustic scenes from TAU Urban Acoustic Scenes 2019 development set \cite{Mesaros2018_DCASE}. Isolated audio samples (i.e. without background noise) from three sound events classes are used to create mixtures of sound events. These mixtures are then used without and with background noise from recordings of vehicle, outdoors, and indoors acoustic scenes, creating four different versions of the same number of sound events mixtures, and under two different signal-to-noise ratios (SNRs). 

\subsection{Processing of sound events and acoustic scenes}
\label{sec:event}
For audio samples of sound events, we use all the available audio files from the TUT Rare Sound Events 2017 development set~\cite{DCASE2017challenge}. We resample each of the audio files to 44.1~kHz and convert them to monophonic. Then, for each audio file in the dataset, we discard the silence period between the consecutive isolated sound events to extract all the 148, 139, and 187 isolated audio samples from ``baby crying'', ``glass breaking'', and ``gunshot'' sound events classes, respectively. This totals to a duration of approximately five, three, and four minutes for the above classes, respectively. The isolated audio samples from each class are split into training, validation, and test such that isolated sound events originating from one audio file in TUT Rare Sound Events 2017 are placed only in one split. These non-overlapping splits of training, validation, and test contain 60\%, 20\%, and 20\% of total amount of isolated sound events for each class, respectively.

As the audio samples of acoustic scenes, we use the development set of the TAU Urban Acoustic Scenes 2019 dataset~\cite{Mesaros2018_DCASE}. This dataset consists of 10 different acoustic scenes which are divided into three categories in the proposed dataset: \textit{vehicle}, \textit{outdoor}, and \textit{indoor}, as presented in Table \ref{tab:table1}. These acoustic scenes have been recorded in different locations of 10 big European cities (e.g. places of interests, neighborhoods) using 48 kHz sampling rate and 24 bits resolution. We resample the audio to 44.1 kHz and 16 bits, to match the sampling rate and bit width of audio samples of sound events. 

\begin{figure}[t!]
    \centering
    \centerline{\includegraphics[width=\linewidth]{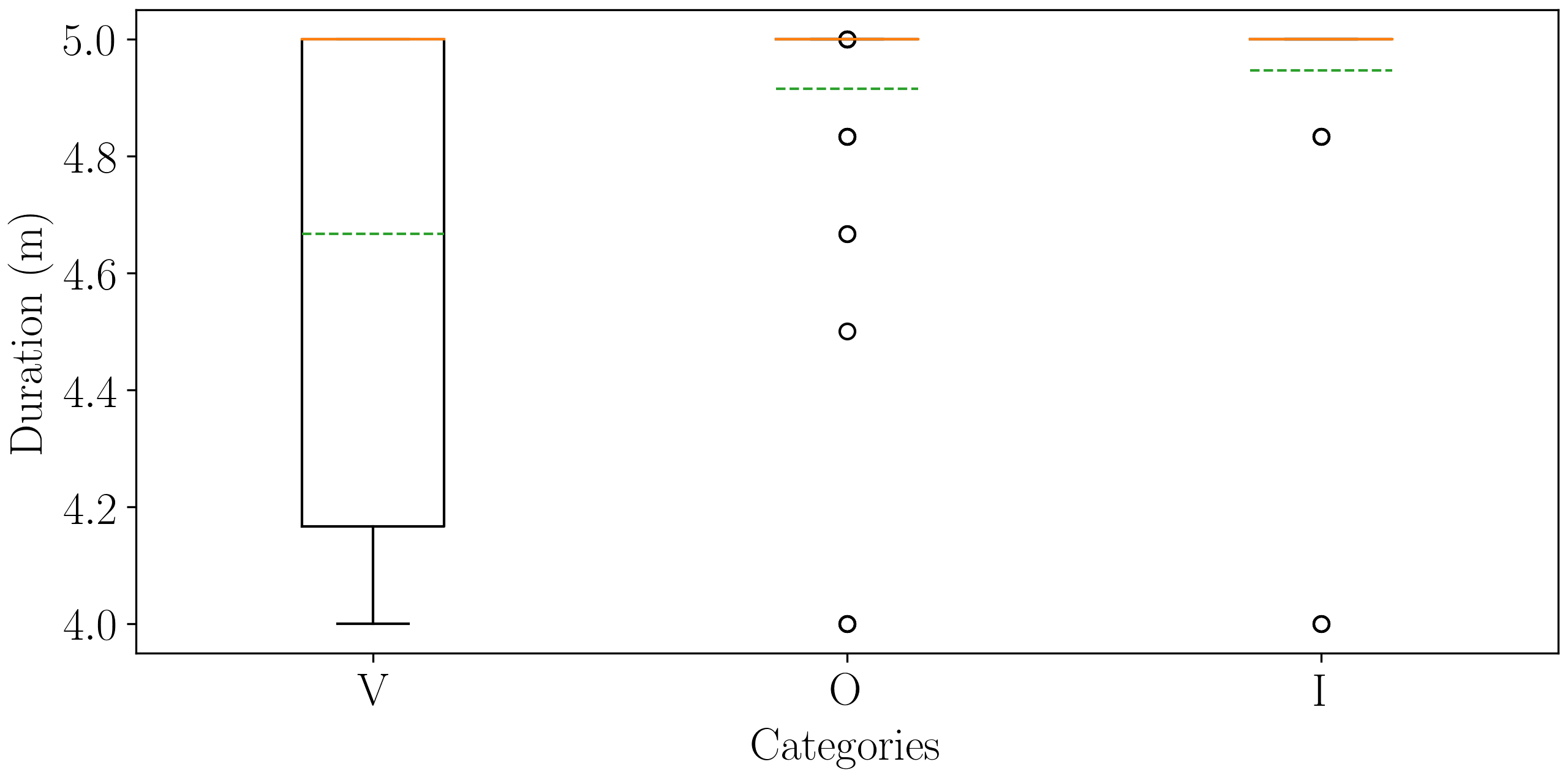}}
    \caption{The distribution of duration for audio samples available from each selected location in different categories of VOICe dataset. The orange lines indicate the median, while the mean is presented by green dashed-lines.}
    \label{fig:as}
\end{figure}

We screen all acoustic scene recordings by listening to find those that contain any audio samples from the sound event classes of ``baby crying'', ``glass breaking'', and ``gunshot''. An audio sample of an acoustic scene is discarded, if it contains any of the aforementioned sound events. In addition, if a location does not have at least 4 minutes of audio, we discard it. In the TAU Urban Acoustic Scenes 2019 development set, the number of locations in outdoor and indoor categories are higher than the number of locations in vehicle. Therefore, we sample the locations from the outdoor and indoor categories to make the number of locations in them the same as in vehicle. As a result, we obtain 69 locations for each category. 
The audio sample of each location was originally published in 10-second segments. We first sort the segments in time, then, we limit the total duration of audio for any selected location between 4 and 5 in minutes by discarding the extra segments. Additionally, we dedicate approximately the first 60\% of the segments to training, the next 20\% to validation, and the remaining 20\% to test. We concatenate the sorted segments of a location in each split and obtain audio samples in the range of 2.4 to 3 minutes for training and 0.8 to 1 minutes for validation and test. These processes allow us to obtain a balance duration of audio samples, in terms of number and minutes of audio, for each location in different categories. Figure \ref{fig:as} illustrates the distribution of duration for audio samples from each location, and separately for each category. 

\subsection{Mixing process}
\label{sec:mix}
\begin{figure}[t!]
    \centering
    \centerline{\includegraphics[width=\linewidth]{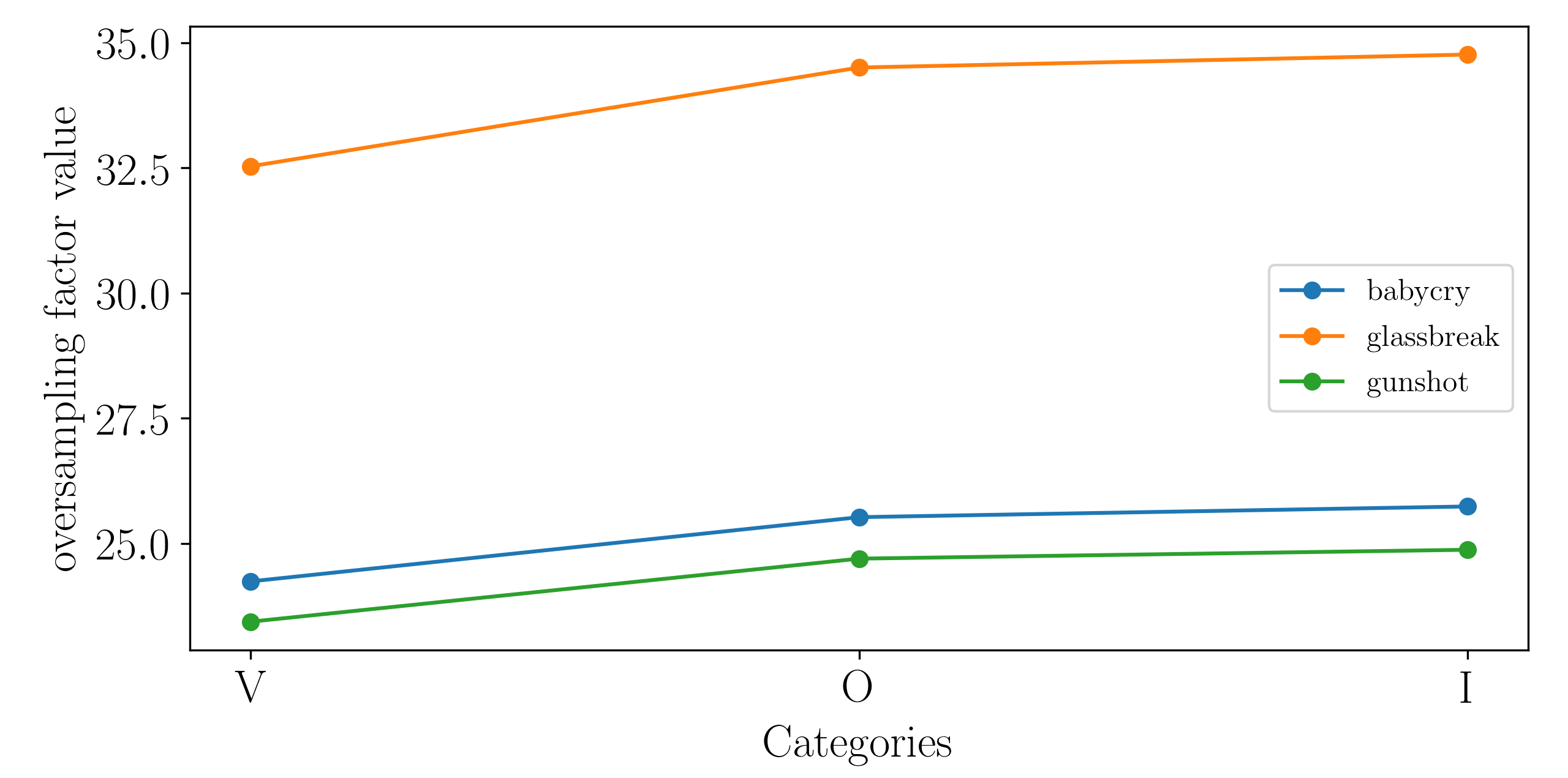}}
    \caption{The average amount of oversampling factor for sound events of created VOICe datasets over two different SNR values.}
    \label{fig:se}
\end{figure}

The VOICe dataset consists of 4 categories: vehicle, outdoor, indoor, and clean (i.e. without background noise). Each category contains 207 audio samples which are the results of 69 locations in three splits (i.e. training, validation, and test). For each category, the process of creating an audio sample is as follows:

An audio sample from a category of acoustic scenes is randomly selected. We synthesize the mixtures of isolated sound events class-by-class. In other words, we create an empty track with the same length as in the selected audio sample of an acoustic scene for each target sound event class. First, a random period of silence is drawn from a discrete uniform distribution (U(1,5) in seconds), and placed at the beginning of the track. Then, we randomly select one isolated sound event of one target class, and place it in the track. This is followed by a random period of silence (U(1,5) in seconds). This process is repeated until the predetermined length of the track is reached. Repeating this process for each target class, we obtain three class-specific tracks, each containing only sound events from one of the target classes. Then, we mix all the three class-specific tracks to create the mixture of all sound events classes. As a result, we have a maximum polyphony of three in our dataset. During this process, once all available isolated sound events from a target class are used, we randomly oversample the isolated sound events for that class using the copies of the same isolated sound events. Figure \ref{fig:se} illustrates the amount of oversampling factor of sound events for the created VOICe datasets with SNR (i.e. signal-to-noise ratio of a sound event mixture to an acoustic scene audio sample) values of -3 and -9 dB on average. 

Finally, we mix the final sound event mixture with the selected audio sample of an acoustic scene to create one audio sample for VOICe dataset, provided that the category is not clean. We repeat the process to create the audio samples of all categories. The length of a created audio sample in VOICe is the same as the length of the selected acoustic scene that we use to mix the sound event mixture with. For the clean category, the lengths of audio samples are chosen such that they are in the same range of length compared to audio samples in the other categories. 

\subsection{Dataset Setup}
\label{sec:setups}

In the context of domain adaptation, the datasets are considered either as source or target domain data. Therefore, if a category (e.g. outdoor) is in the role of source domain in our experiment, we keep its original training, validation, and test splits to address SED task properly. However, if it is in the role of target domain data, the data from training and validation splits are mixed into a single split named adaptation. This results in two splits in total for target domain data: adaptation and test. The daptation split is used during the actual adaptation process, while the evaluation of the adaptation process should be reported on test split. Moreover, we need to first tune the hyper-parameters of an adaptation method for a target domain dataset. Then, using the tuned parameters, we are able to apply the method for a different target domain dataset, and to evaluate the performance of the method in generalized manner. In this regard, four different categories of VOICe dataset enable us to perform a generalized domain adaptation using different categories in different roles of source and target datasets.

\section{Experiments setup}
\label{sec:baseline}
We demonstrate our results using two different settings. Our first setting evaluates the adaptation performance among vehicle, outdoor, and indoor categories. The second setting aims at evaluating the performance of an adaptation method from the clean category as a source dataset to categories with background noise (e.g. vehicle) as target datasets.

\subsection{Detection and adaptation tasks}
\label{ssec:sed}
The adaptation method of our work was initially introduced in~\cite{Gharib2018}. This method is implemented in three main steps: detection task (i.e. SED), adaptation, and evaluation. 
Firstly, we perform the detection task (i.e. SED) using the training, validation, and test splits of source domain dataset. For this work, we employ a CRNN architecture which was initially introduced in~\cite{7933050} as our model and sound event detector. The model contains three blocks of convolutional neural networks (CNNs), Each block uses a convolutional layer, batch normalization, ReLU activation, max pooling, and dropout in the mentioned order. The sound event detector consists of a layer of gated recurrent units (GRU) in addition to a feedforward neural network (FNN) as the output layer. As a result of this step, we obtain a model (i.e. source model) together with a sound event detector trained on source domain dataset. The input to our model is 40 log-mel band energies. These features are extracted using hamming windows of 1024 samples (i.e. $\sim$ 23 ms) with 50\% overlap.

In the second step, we aim at adapting the source model to a target domain dataset in an unsupervised manner. We create a new model using the same architecture as in the source model, and initialize its weights using the weights of the source model. This model is named target model. Moreover, we also employ a domain classifier whose architecture consists of two layers of FNNs. The input to the source model is the source domain data while the inputs to the target model are from both domains. The domain classifier receives the latent representations of the source and target domain data from source and target model respectively to distinguish the domain of its inputs. In addition, the latent representations of the source domain data from the target model are given to the sound event detector to also update the target model in order to not forget the learned information of the source domain data while being adapted to the target domain dataset. We place the domain classifier and the target model in a competition against each other using an adversarial training scheme in order for the target model to learn an invariant representation of source and target domain data to fool the domain classifier in believing that the latent representations of the target domain data are the same as that of the source domain data. Eventually, after the adaptation process, the target model and sound event detector are used to perform SED task on target domain data.

\subsection{Results}
\label{ssec:results}
Table \ref{tab:table2} presents the results of SED and adaptation tasks for VOICe dataset. In order to eliminate the random effects, we have run each of the experiments three times. The results are shown using the average value together with standard deviation. Table~\ref{tab:table2} is an example of how to use the categories of VOICe dataset in different roles as source (i.e. $\mathcal{D}_S$) and target (i.e. $\mathcal{D}_T$) domain data. Each row of Table~\ref{tab:table2} presents an adaptation process from $\mathcal{D}_S$ to $\mathcal{D}_T$. In addition, $\boldsymbol{F_{1_S}^{S_{te}}}$ and $\boldsymbol{F_{1_S}^{T_{te}}}$ show the $F_1$-scores of source model for test split of $\mathcal{D}_S$ and $\mathcal{D}_T$ respectively, while $\boldsymbol{F_{1_A}^{T_{te}}}$ shows the $F_1$-score of adapted model (i.e. target model) on test split data from $\mathcal{D}_T$. Hyper-parameters of adaptation process are selected randomly. Therefore, we have not tuned them for the results in $\boldsymbol{F_{1_A}^{T_{te}}}$. Based on the $\boldsymbol{F_{1_S}^{S_{te}}}$ scores, the performance is clearly dropped on all categories except for three experiments namely I / V, I / O, and O / V. This can be explained by comparing $\boldsymbol{F_{1_S}^{S_{te}}}$ scores in the table. Among the categories with background noise, $\boldsymbol{F_{1_S}^{S_{te}}}$ scores for the vehicle category is higher than the outdoor category. Furthermore, the outdoor category performs better than the indoor category. These results indicate the indoor as the most difficult and the vehicle as the easiest category of acoustic scenes in VOICe dataset for SED task. This explains why the model which is optimized on the indoor category (as $\mathcal{D}_S$) performs better on the outdoor and vehicle (as $\mathcal{D}_T$). The same reason can be also applied for the O / V experiment. Lastly, the clean category exhibits the highest $\boldsymbol{F_{1_S}^{S_{te}}}$ scores compared to other categories due to no background noise. However, the results drastically drop when it is evaluated using categories with background noise.

\begin{table}[t!]
\centering
\caption{$F_1$-score of SED task for source and target models. V., O., I., Ce. stand for vehicle, outdoor, indoor, and clean categories respectively.}
\label{tab:table2}
\bigskip
\resizebox{\columnwidth}{!}{%
\begin{tabular}{l|ccc}
\multicolumn{1}{l}{} & \multicolumn{2}{r}{\textbf{-9 dB SNR}} \\ \cline{2-4}
\addlinespace[3pt]
\multicolumn{1}{l}{$\boldsymbol{\mathcal{D}_S} / \boldsymbol{\mathcal{D}_T}$} & $\boldsymbol{F_{1_S}^{S_{te}}}$ & $\boldsymbol{F_{1_S}^{T_{te}}}$ & $\boldsymbol{F_{1_A}^{T_{te}}}$ \\ \hline
V. / O. & 0.80$\pm$0.003 & 0.73$\pm$0.003 & 0.74$\pm$0.006\\
V. / I. & 0.81$\pm$0.008 & 0.69$\pm$0.002 & 0.66$\pm$0.018\\ \hline
O. / V. & 0.75$\pm$0.010 & 0.77$\pm$0.007 & 0.76$\pm$0.023\\
O. / I. & 0.75$\pm$0.008 & 0.70$\pm$0.011 & 0.69$\pm$0.013\\ \hline
I. / V. & 0.73$\pm$0.002 & 0.75$\pm$0.008 & 0.73$\pm$0.007\\ 
I. / O. & 0.74$\pm$0.002 & 0.75$\pm$0.008 & 0.70$\pm$0.033\\ \hline
Ce. / V. & 0.92$\pm$0.003 & 0.50$\pm$0.017 & 0.43$\pm$0.011\\
Ce. / O. & 0.92$\pm$0.003 & 0.49$\pm$0.022 & 0.42$\pm$0.010\\
Ce. / I. & 0.91$\pm$0.004 & 0.50$\pm$0.007 & 0.40$\pm$0.095\\
\end{tabular}
}
\end{table}

\section{Conclusion}
\label{sec:conclusion}
In this paper, we emphasized on the importance of generalizable domain adaptation in order to access the advantages and limits of adaptation methods in realistic scenarios of SED task. Moreover, we introduced our first SED dataset for the purpose of generalized domain adaptation, i.e. VOICe. The three categories of VOICe: vehicle, outdoor, and indoor can be used to simulate a realistic scenario in which there is a mismatch of acoustic scene conditions between audio recordings of data on which a model is optimized for and the data we evaluate the model on. In addition, VOICe also contains a category without background noise which is called clean. We also presented the baseline method and the evaluation results of SED and domain adaptation for VOICe dataset.

\section{Acknowledgement}
\label{sec:acknowledgement}
The research leading to these results has received funding from the European Research Council under the European Union’s H2020 Framework Programme through ERC Grant Agreement 637422 EVERYSOUND.
Part of the computations leading to these results were performed on a TITAN-X GPU donated by NVIDIA to K. Drossos. The authors wish to acknowledge CSC–IT Center for Science in Finland for computational resources.

\bibliographystyle{IEEEbib}
\bibliography{strings,refs}

\end{document}